\documentclass[aps,letterpaper,nofootinbib,preprint,showpacs,amsmath,amssymb,pdftex11pt]{revtex4-2}%
\usepackage{latexsym}

\usepackage{latexsym}%

\usepackage{graphicx} % Required for inserting images

\usepackage{color}

\newcommand{\be}{\begin{equation}}
\newcommand{\ee}{\end{equation}}

\newcommand{\cG}{\mathcal{G}}
\newcommand{\cL}{\mathcal{L}}

\allowdisplaybreaks

\begin{document}
%---------------------------------------------------------------------
\begin{center}
{\bf{\Large Rethinking the Effective Field Theory \\  formulation of Gravity}}  \\
\bigskip
Essay written for the Gravity Research Foundation 2024 \\ Awards for Essays on Gravitation.
\bigskip

March 29, 2024
\bigskip

\renewcommand{\thefootnote}{\fnsymbol{footnote}}
Jesse Daas\footnote{J.Daas@science.ru.nl}, Cristobal Laporte\footnote{cristobal.laportemunoz@ru.nl}$^{,}$\footnote{Corresponding Author}, Frank Saueressig\footnote{f.saueressig@science.ru.nl}, and Tim van Dijk\footnote{tim.vandijk@ru.nl}
\smallskip
\renewcommand{\thefootnote}{\arabic{footnote}}

{\it Institute for Mathematics, Astrophysics, and Particle Physics,}\\
{\it Radboud University, Nijmegen, The Netherlands}\\
\medskip

\end{center}
\medskip

%---------------- Abstract here ----------------------------
\noindent
General relativity is highly successful in explaining a wide range of gravitational phenomena including the gravitational waves emitted by binary systems and the shadows cast by supermassive black holes. From a modern perspective the theory is not fundamental though, but constitutes the lowest order term in an effective field theory description of the gravitational force. As a consequence, the gravitational dynamics should receive corrections by higher-derivative terms. This essay discusses structural aspects associated with these corrections and summarizes their imprint on static, spherically symmetric geometries. Along these lines, we critically reassess the common practice of using local field redefinitions in order to simplify the dynamics at the danger of shifting physics effects into sectors which are beyond the approximation under consideration.
%---------------------------------------------------------------------

\maketitle
%---------------------------------------------------------------------
\section{Introduction}
%---------------------------------------------------------------------
General relativity provides the gold standard when it comes to predicting phenomena related to gravity \cite{Will:2014kxa}. During the last decade, tests related to the direct detection of gravitational waves \cite{LIGOScientific:2016aoc,KAGRA:2021duu} and images showing the shadows cast by supermassive black holes \cite{EventHorizonTelescope:2019dse,EventHorizonTelescope:2019ggy,EventHorizonTelescope:2022wkp} have further corroborated this picture. These new classes of tests rely on sophisticated numerical simulations which are then compared to data. This step makes manifest use of a predetermined spacetime geometry with the solutions of general relativity, given by the Schwarzschild solution and its extension including spin, constituting the canonical choice. 
This approach leads to a perfect agreement between theoretical predictions and observations. At the same time, it raises the question whether there are generic properties of spacetime, as, e.g., naked singularities, which are ruled out
 by the observations. Phrased differently, one may ask whether the spacetimes arising from general relativity are ``generic'', e.g., in terms of their horizon structure and no-hair theorems. This calls for a systematic understanding of the gravitational fields created by compact isolated objects beyond the dynamics dictated by general relativity.

Theoretical guidance towards answering such a question is expected to arise, e.g., from a quantum theory of the gravitational force. While there are several well-established proposals for such a theory, including string theory and the holographic principle \cite{Zwiebach:2004tj,Becker:2006dvp}, loop quantum gravity \cite{Rovelli:2014ssa,Ashtekar:2017yom}, and the gravitational asymptotic safety program \cite{Percacci:2017fkn,Reuter:2019byg}, there is no consensus yet. In this light, one may resort to effective field theory techniques to address this question. This approach starts from the Einstein-Hilbert action and supplements the dynamics by higher-derivative terms coming with free, unknown couplings. These corrections may have their origin either in the quantum theory or arise from integrating out heavy degrees of freedom. Being agnostic about their origin, this strategy allows to study the impact of these corrections on the spacetime geometry in a model-independent way.

In this essay, we follow this strategy and systematically review the corrections related to local effective interactions containing up to six derivatives. Anticipating that ultimately the pure gravity theory will be supplemented by matter degrees of freedom, we refrain from simplifying the analysis by eliminating potential interaction terms through a field redefinition of the spacetime metric. In this way, we identify the parameterized post-Newtonian (PPN) order at which a given higher-derivative term starts to contribute also at the non-linear level. This investigation is closely related to recent efforts to construct black hole spacetimes in the framework of effective field theory \cite{Cardoso:2018ptl,Cano:2019ore,deRham:2020ejn,Cano:2020cao}, quadratic gravity \cite{Lu:2015cqa,Lu:2015psa,Svarc:2018coe,Daas:2022iid}, Einstein-Cubic gravity \cite{Hennigar:2016gkm,Bueno:2016lrh,Adair:2020vso,Gutierrez-Cano:2024oon}, and in the presence of the Goroff-Sagnotti counterterm \cite{Alvarez:2023gfg,Daas:2023axu}.

%---------------------------------------------------------------------
\section{The derivative expansion of the gravitational action}
%---------------------------------------------------------------------
A key feature of general relativity is that its equations of motion transform covariantly with respect to coordinate transformations. Combining this idea with the requirement that the admissible spacetimes are stationary points of an action functional (containing at most two derivatives of the spacetime metric $g_{\mu\nu}$), one directly arrives at the Einstein-Hilbert (EH) action\footnote{The relevance of being able to derive alternatives to the spacetimes appearing within general relativity based on action principles has recently been stressed in \cite{Knorr:2022kqp}.}
\be\label{EH-action}
S^{\rm EH}[g] = \frac{M_{\rm Pl}^2}{16 \pi} \int d^4x \sqrt{-g} R \, . 
\ee
Here $g$ is the determinant of the spacetime metric, $R$ denotes its Ricci scalar and we have traded Newton's constant $G$ for the Planck mass $M_{\rm Pl}^2 \equiv G^{-1}$. For the sake of keeping the discussion simple, we neglect the cosmological constant and do not display the contributions from matter degrees of freedom. The stationary geometries arising from \eqref{EH-action} then satisfy
\be\label{EOM1}
R_{\mu\nu} - \frac{1}{2} g_{\mu\nu} R = 0 \, ,
\ee
where $R_{\mu\nu}$ is the Ricci tensor. In the absence of matter, eq.\ \eqref{EOM1} simplifies to Einstein's field equations in vacuum $R_{\mu\nu} = 0$. The fact that \eqref{EOM1} has been derived from an action which is invariant with respect to coordinate transformations automatically guarantees the conservation of the stress-energy tensor based on the symmetries of the theory. 

%---------------------------------------------------------------------
\subsection{Organizing the expansion}
%---------------------------------------------------------------------
%The effective field theory viewpoint implies that the Einstein-Hilbert action receives corrections either through quantum effects or by taking into account the dynamics of massive degrees of freedom at an effective level. 
In the weak field regime, one expects that the corrections to the Einstein-Hilbert action can be captured by a derivative expansion of the gravitational dynamics 
\be\label{Dn-exp}
S^{\rm der}[g] = \frac{1}{16\pi} \int d^4x \sqrt{-g} \left[ M_{\rm Pl}^4  \cL_{\rm D0} + M_{\rm Pl}^2 \cL_{\rm D2} + \cL_{\rm D4} + \frac{1}{M_{\rm Pl}^2} \cL_{\rm D6}+ \frac{1}{M_{\rm Pl}^4}  \cL_{\rm D8}+ \cdots\right] \, . 
\ee
Here the subscript ${\rm Dn}$ indicates that the terms contain $n$ derivatives and we singled out the Planck mass $M_{\rm Pl}$ in order to set the ``natural'' size of the higher-order contributions. Again neglecting the cosmological constant, $\cL_{\rm D0} \approx 0$, the Einstein-Hilbert action, $\cL_{\rm D2} = R$, is the leading term in this expansion. Depending on the situation, this scale may be replaced by a different mass-scale, e.g., the one associated with the heavy physics generating the effective dynamics \cite{Dvali:2007hz}.

It is then natural to inquire about the structure of the subleading terms. Canonically, one imposes that the building blocks making up the $\cL_{\rm Dn}$, transform as scalars with respect to general coordinate transformations and are local. For the sake of simplicity, we are also restricting to contributions which are parity even. Following the classification \cite{Fulling:1992vm}, the $\cL_{\rm Dn}$ can be constructed from contractions of the Riemann tensor $R_{\mu\nu\rho\sigma}$ and covariant derivatives $\nabla_\mu$ acting on it.
Owed to the Bianchi-identities and symmetries of the Riemann tensor, there are various options for writing the set of independent interaction terms, see, e.g., \cite{Decanini:2007gj,Cano:2019ore,deRham:2020ejn,Cano:2021myl} and  \cite{Decanini:2008pr} for a collection of identities relating these choices. From the physics perspective, it is sometimes more convenient to trade the Riemann tensor for the Weyl tensor $C_{\mu\nu\rho\sigma}$, so that the curvature invariants are generated by the set $\{R, R_{\mu\nu}, C_{\mu\nu\rho\sigma}, \nabla_\mu\}$.  

In this essay, we adopt the following choice. At order $n=4$ in the derivative expansion we use
\be
\cL_{\rm D4} = \beta \, R^2 - \alpha \, C_{\mu\nu\rho\sigma} \, C^{\mu\nu\rho\sigma} + \gamma \, \mathcal{G}
\ee
where $\cG \equiv  R^2 - 4 R_{\mu\nu} R^{\mu\nu} + R_{\mu\nu\rho\sigma} R^{\mu\nu\rho\sigma} $ is the integrand of the Gauss-Bonnet topological invariant. At order $n=6$ there are ten independent invariants. These are conveniently organized according to their physics implications,
\be\label{D6class}
\cL_{\rm D6} = \cL_{\rm D6}^{\rm GS} + \cL_{\rm D6}^{\rm kin} + \cL_{\rm D6}^{\rm geo} + \cL_{\rm D6}^{\rm field} \, .
\ee
Setting $\square \equiv \nabla^\mu \nabla_\mu$, we take
\be\label{D6class2}
\begin{split}
\cL_{\rm D6}^{\rm kin} = & \, \eta_1 \, R \, \square \, R - \eta_2 \, C_{\mu\nu\rho\sigma} \, \square \, C^{\mu\nu\rho\sigma} \, , \\
\cL_{\rm D6}^{\rm field} = & \, \eta_3 \, R^3 + \eta_4 \, R \, R_{\mu\nu} \, R^{\mu\nu} + \eta_5 \, R \, C_{\mu\nu\rho\sigma} \, C^{\mu\nu\rho\sigma} 
     + \eta_6 \, R^{\nu}_{\mu} \, R^{\rho}_{\nu} \, R^{\mu}_{\rho} + \eta_7 \, R^{\mu\rho} \, R^{\nu\sigma} \, C_{\mu\nu\rho\sigma} \, ,
     \\
\cL_{\rm D6}^{\rm geo} = & \, \eta_8 \, R^{\sigma\delta}\, C_{\mu\nu\rho\sigma} \, C^{\mu\nu\rho}{}_{\delta} + \eta_{10} \, C^{\phantom{\mu}\rho\phantom{\nu}\sigma}_{\mu\phantom{\rho}\nu} \, C^{\phantom{\rho}\delta\phantom{\sigma}\gamma}_{\rho\phantom{\delta}\sigma} \, C^{\phantom{\delta}\mu\phantom{\gamma}\nu}_{\delta\phantom{\mu}\gamma} \, , \\
\cL_{\rm D6}^{\rm GS} = & \, \eta_9 \, C^{\phantom{\mu\nu}\rho\sigma}_{\mu\nu} \, C^{\phantom{\rho\sigma}\delta\gamma}_{\rho\sigma} \, C^{\phantom{\delta\gamma}\mu\nu}_{\delta\gamma} \, .
\end{split}
\ee
The rationale underlying this grouping is the following: terms that can be eliminated by geometric identities specific to four spacetime dimensions are collected in $\cL_{\rm D6}^{\rm geo}$. The terms in $\cL_{\rm D6}^{\rm kin}$ and $\cL_{\rm D6}^{\rm field}$ can be eliminated by a linear field redefinition at the level of the action. The contributions $\cL_{\rm D6}^{\rm kin}$ are thereby singled out since they also contribute to the graviton propagator in a flat spacetime. In this sense, these terms are part of the gravitational form factors determining the graviton propagator in this background \cite{Knorr:2019atm,Knorr:2022dsx}. In an effective field theory setting, these may be eliminated by a field redefinition in a consistent way. Finally, $\cL_{\rm D6}^{\rm GS}$, corresponds to the Goroff-Sagnotti (GS) counterterm appearing in the quantization of general relativity \cite{Goroff:1985th} and seeds the corrections to the dynamics entailed by general relativity. The contributions at higher orders in the derivative expansion can be grouped in a similar way. For the purpose of this work, we limit ourselves to one term
\be\label{D8example}
\cL_{\rm D8} = \rho_1 (R_{\mu\nu\rho\sigma} \, R^{\mu\nu\rho\sigma})^2 + \cdots \, . 
\ee
The latter again generates a non-trivial correction to the spacetimes found in general relativity, and serves as a benchmark for the order of the weak-field expansion at which these corrections appear.
%---------------------------------------------------------------------
\subsection{Simplifying the dynamics}
\label{sect.2.B}
%---------------------------------------------------------------------
Starting from \eqref{D6class}, there are two ways which allow to simplify the dynamics arising from the derivative expansion: geometrical identities and field redefinitions. 

\noindent
\emph{Simplifications due to geometrical identities.} Working in four spacetime dimensions is special in various ways. Firstly, the Gauss-Bonnet term $\cG$ integrates to a topological invariant. As a result, it is independent of the spacetime metric and does not enter into the equations of motion. This allows to eliminate either the contribution of $R_{\mu\nu}R^{\mu\nu}$ or, alternatively, the Weyl-squared term from $\cL_{\rm D4}$. Among the terms contained in $\cL_{\rm D6}$, one may exploit that in four spacetime dimensions the anti-symmetrization of five or more spacetime indices has to vanish. This leads to the relations
\be\label{geo-ids}
\begin{split}     
   C^{\phantom{\gamma}\rho\phantom{\mu}\sigma}_{\gamma\phantom{\rho}\mu} \, C^{\phantom{\sigma}\nu\phantom{\rho}\delta}_{\sigma\phantom{\nu}\rho} \, C^{\phantom{\delta}\gamma\phantom{\nu}\mu}_{\delta\phantom{\gamma}\nu} = & \, \frac{1}{8} R \, C_{\gamma\delta\mu\nu} \, C^{\gamma\delta\mu\nu} + \frac{1}{2} C^{\phantom{\mu\nu}\rho\sigma}_{\mu\nu} \, C^{\phantom{\rho\sigma}\delta\gamma}_{\rho\sigma} \, C^{\phantom{\delta\gamma}\mu\nu}_{\delta\gamma} \, , \\
    R^{\delta\mu} C_{\delta}{}^{\nu\rho\sigma} \, C_{\mu\nu\rho\sigma} = & \frac{1}{4} R \, C_{\delta\mu\nu\rho} \, C^{\delta\mu\nu\rho} ,
\end{split}
\ee
These allow to express the terms contained in $\cL_{\rm D6}^{\rm geo}$ in terms of the other curvature scalars. Note that these identities are purely geometrical and can always be evoked.

\noindent
\emph{Simplifications due to field redefinitions.} In addition, \eqref{Dn-exp} is commonly simplified by a field redefinition of the metric \cite{tHooft:1974toh,Anselmi:2013wha,deRham:2020ejn,Knorr:2023usb,Wetterich:2024uub},
\be\label{fieldredef}
g_{\mu\nu} \mapsto g_{\mu\nu} + \delta g_{\mu\nu} = g_{\mu\nu} - \frac{16 \pi}{M_{\rm Pl}^2}\big[a_1 \,R_{\mu\nu} + a_2\, g_{\mu\nu}R\big] + \cdots \, , 
\ee
where the $\cdots$ symbolize terms with at least four spacetime derivatives. Starting with $\cL_{\rm D2}$ and working at linear order yields a change in the action proportional to the equations of motion \eqref{EOM1}
\be\label{fieldredef2}
\delta S^{\rm EH}[g] = \frac{M_{\rm Pl}^2}{16 \pi} \int d^4x \sqrt{-g} \left[R^{\mu\nu} - \frac{1}{2} g^{\mu\nu}R \right] \delta g_{\mu\nu} \, . 
\ee
Substituting $\delta g_{\mu\nu}$ from \eqref{fieldredef} and choosing $ a_1 = 4\alpha$, $a_2 = -\frac{2}{3}\alpha + 2\beta$ allows to eliminate $\cL_{\rm D4}$, provided that the Weyl-squared term is removed by the Gauss-Bonnet term. Along the same lines, higher-derivative terms in \eqref{fieldredef} allow to eliminate $\cL_{\rm D6}^{\rm kin}$ and $\cL_{\rm D6}^{\rm field}$. Thus, after applying the geometric identities and field redefinitions, the only term left at order $n=6$ is $\cL_{\rm D6}^{\rm GS}$ \cite{Goroff:1985th}.

%---------------------------------------------------------------------
\section{Static spacetimes exhibiting spherical symmetry}
%---------------------------------------------------------------------
At this point, it is interesting to discuss the impact of the higher-derivative terms on static, spherically symmetric spacetime geometries. The most general line-element compatible with these symmetries can be cast into the form
\be\label{lineansatz}
  ds^2 = -h(r) \, dt^2 + \frac{1}{f(r)} \, dr^2 + r^2 \left(d\theta^2 + \sin{\theta} \, d\phi^2\right) \, . 
\ee
Thus, we are dealing with two free metric functions $h(r), f(r)$ which determine the geometry. Substituting this ansatz into \eqref{EOM1} yields the Schwarzschild solution
\be\label{SS-geo}
h(r) = f(r) = 1 - \frac{2 G M}{r} \, . 
\ee
The integration constant $M$ determines the asymptotic mass of the solution. This case is special in the sense that the two metric functions are equal. The goal of this section is to discuss the modifications of this solution triggered by the derivative expansion.
%---------------------------------------------------------------------
\subsection{Linearized solutions in the asymptotically flat regime}
%---------------------------------------------------------------------
Our first step consists in exhibiting the additional degrees of freedom provided by the higher-derivative terms. This is readily done by deriving the equations of motion resulting from \eqref{Dn-exp} and studying linear perturbations around flat space
\be\label{linear-pert}
h(r) = 1 + \varepsilon \, h_{\rm lin}(r) \, , \qquad 
f(r) = 1 + \varepsilon \, f_{\rm lin}(r) \, . 
\ee
Working at first order in $\varepsilon$, only terms with at most two powers of the spacetime curvature contribute to the dynamics. This entails that the linearized equations of motion receive contributions from $\cL_{\rm D2}$, $\cL_{\rm D4}$, and $\cL_{\rm D6}^{\rm kin}$ only. The solution of the resulting linear system of differential equations determining $h_{\rm lin}(r)$ and $f_{\rm lin}(r)$ are readily found in terms of linear combinations of exponentials. Setting $G=1$ for the sake of conciseness, one has\footnote{This result is readily extended to include the kinetic terms $\cL_{\rm Dn}^{\rm kin}$ up to order $n$. Each order introduces a new pair of integration constants $\{A^+,A^-\}$ for each sector $C,R$ and the sums extend from $i=1,\cdots,n/2-1$.}
\begin{align}\label{linsol}
%\begin{split}
    h_{\rm lin}(r) = & \, C_t -\frac{2M}{r}   \, + \sum_{i=1}^2 \left[ \frac{A^{C-}_i}{r}\,e^{-m^{C}_i\,r} -2\,\frac{A^{R-}_i}{r}\,e^{-m^{R}_i\,r} +
    \frac{A^{C+}_i}{r}\,e^{m^{C}_i\,r} -2\,\frac{A^{R+}_i}{r}\,e^{m^{R}_i\,r}
    \right],\\
    f_{\rm lin}(r) = & \, -\frac{2M}{r} + \sum_{i=1}^2 \left[ \frac{A^{C-}_i}{2r}\big(1 + \,m^{C}_i\,r\big)\,e^{-m^{C}_i\,r} + \frac{A^{R-}_i}{2r}\big(1 + \,m^{R}_i\,r\big)\,e^{-m^{R}_i\,r} \right. \nonumber\\
    & \qquad \qquad \qquad \quad \left. + \frac{A^{C+}_i}{2r}\big(1 - \,m^{C}_i\,r\big)\,e^{m^{C}_i\,r} + \frac{A^{R+}_i}{2r}\big(1 - \,m^{R}_i\,r\big)\,e^{m^{R}_i\,r} \right] \, . 
%\end{split}
\end{align}
The contributions from the Weyl sector $(C)$ come with the masses
\begin{equation}
    \big(m^{C}_{1}\big)^2 = \frac{-\alpha + \sqrt{\alpha^2 + 2\eta_2}}{2\eta_2}\,,\qquad \big(m^{C}_{2}\big)^2 = \frac{-\alpha - \sqrt{\alpha^2 + 2\eta_2}}{2\eta_2} \,,
\end{equation}
while the terms containing the Ricci scalar $(R)$ are associated with
\begin{equation}
    \big(m^{R}_{1}\big)^2 = \frac{-3\beta + \sqrt{9\beta^2 + 6\eta_1}}{6\eta_1} \, ,\qquad \big(m^{R}_2\big)^2 = \frac{-3\beta - \sqrt{9\beta^2 + 6\eta_1}}{6\eta_1}\, .
\end{equation}
For $\alpha, \beta > 0$, the masses $m_i^C$ and $m_i^R$ are real if $-\frac{\alpha^2}{2} < \eta_2 \leq 0$ and $-\frac{3\beta^2}{2} < \eta_1 \leq 0$, respectively. The masses associated with the massive degrees of freedom familiar from quadratic gravity \cite{Stelle:1977ry} are recovered from $m_1^C$ and $m_1^R$ in the limit $\eta_1, \eta_2 \rightarrow 0$.

The linearized solution \eqref{linsol} has 10 integration constants $\{C_t, M, A_i^{C-}, A_i^{C+}, A_i^{R-}, A_i^{R+}\}$. Structurally, the massless sector contributes $\{ C_t, M \}$ while each massive mode generated by the higher-derivative terms gives rise to Yukawa-type contributions with two free parameters $\{A^+,A^-\}$. Imposing a canonical normalization of the time-coordinate fixes $C_t = 0$. Moreover, asymptotic flatness dictates that the integration constants associated with exponentially growing Yukawa contributions $\{A_i^{C+},A_i^{R+}\}$ should be set to zero. Thus, working up to order $n=6$ in the derivative expansion allows for asymptotically flat solutions characterized by 5 free parameters.

At this stage, several comments are in order. From the perspective of an effective field theory, the massive degrees of freedom are considered as beyond the validity of the approximation, since the effective theory should contain degrees of freedom of general relativity only. From the perspective of a quantum theory, the massive degrees of freedom can be responsible for rendering the theory (super-)renormalizable \cite{Stelle:1976gc,Asorey:1996hz}. 
%The prototypical example is quadratic gravity \cite{Stelle:1976gc}, where the role of the ghost degrees of freedom has recently been discussed extensively in \cite{Anselmi:2018tmf,Donoghue:2019ecz}. 
Structurally, $\cL_{\rm D4}$ and $\cL_{\rm D6}^{\rm kin}$ are part of the  gravitational form factors determining the graviton propagator in flat space \cite{Knorr:2019atm,Knorr:2022dsx}. From this perspective, the Yukawa-type contributions could arise as an artifact of approximating the exact form factor within the derivative expansion \cite{Platania:2020knd,Platania:2022gtt}. Thus, the interpretation of \eqref{linsol} varies on a case to case basis and needs to be considered with care.

%---------------------------------------------------------------------
\subsection{Power-law corrections to the Schwarzschild geometry}
%---------------------------------------------------------------------
Finally, we determine the power-law corrections to the metric functions \eqref{SS-geo} generated by the higher-derivative terms. In variance with common practice, we simplify the $\cL_{\rm Dn}$ using geometric identities only and refrain from cutting down the remaining terms by field redefinitions. As a consequence, all corrections are imprinted in the geometry of spacetime which then provides the universal stage for the matter fields.

The power-law corrections are readily found via the Frobenius method. The metric functions are expanded in terms of inverse powers of $r$,
\be\label{metexp}
h(r) = 1 + \sum_{n=1} \frac{a_n}{r^n} \, , \qquad f(r) = 1 + \sum_{n=1} \frac{\tilde{a}_n}{r^n} \, , 
\ee
where the leading term has been fixed by imposing asymptotic flatness. The unknown series coefficients $\{a_n, \tilde{a}_n\}$ are determined recursively by substituting the expansion into the equations of motion and solving the resulting hierarchy of equations order by order in $1/r$. Abbreviating $r_s \equiv 2 G M$ and displaying terms up to order $\mathcal{O}(r^{-9})$, this gives
\be\label{fhexp}
\begin{split}
    h(r) \simeq \, &  1 - \frac{r_s}{r} + \frac{6  \, \left[4 \, \eta_2 - 2 \eta_5 + \eta_9 \right] \, r_s^2}{M_{\rm Pl}^4 \, r^6} - \frac{\left[21 \eta_2 - 18 \eta_5 + 4  \eta_9\right] \, r_s^3}{M_{\rm Pl}^4 \, r^7} \\
    &+ \frac{360   \left[ 3 \, \beta \, (\eta_2 - 2 \eta_5) + \alpha \, (3 \, \eta_2 + \eta_9)\right] \, r_s^2}{M_{\rm Pl}^6 \, r^8}\\
    & + \frac{  4 \, \left[32 \rho_1 - 163 \alpha \left(3 \eta_2 + \eta_9\right) - 3 \beta \left(249 \eta_2 - 486 \eta_5 + 2 \eta_9\right)\right] \, r_s^3}{M_{\rm Pl}^6 \, r^9} \, , 
  \\
    f(r) \simeq \, &1 - \frac{r_s}{r} + \frac{18 \, \left[\eta_2 + 4 \, \eta_5 + \eta_9\right] r_s^2}{M_{\rm Pl}^4 \, r^6} - \frac{ \left[15 \eta_2 + 66 \eta_5 + 16 \eta_9\right] \, r_s^3}{M_{\rm Pl}^4 \, r^7} \\
    &+ \frac{1440  \left[\alpha \left(\eta_9 + 3 \eta_2\right) - 6 \beta \left(\eta_2 - 2 \eta_5\right)  \right] \, r_s^2}{M_{\rm Pl}^6 \, r^8} \\
    &+ \frac{36  \left[16 \rho_1 + 3 \beta \left(189 \eta_2 - 366 \eta_5 + 2 \eta_9\right) - 89 \alpha \left(\eta_9 + 3 \eta_2\right) \right] \, r_s^3}{M_{\rm Pl}^6 \, r^9} \, . 
\end{split}
\ee

This result exhibits several interesting properties. First, keeping the couplings fixed there is only one free parameter $M$. In particular, the integration constants associated with the Yukawa terms are absent. This is readily understood from the fact that these corrections are non-analytic at the expansion point and therefore missed by the ansatz \eqref{metexp} \cite{Saueressig:2021wam}. Second, the leading corrections to \eqref{SS-geo} originate from $\cL_{\rm D6}$ and appear at the 6$^{\rm th}$ order in the PPN-expansion. This reflects that the Schwarzschild geometry is also a solution of the equations of motion obtained at order $\cL_{\rm D4}$ (quadratic gravity). The couplings from $\cL_{\rm D4}$ first appear at 8$^{\rm th}$ order. Their structure indicates that the $\cL_{\rm D6}$-terms are needed as a source for these corrections. Third, the benchmark term $\cL_{\rm D8}$ first appears at order $r^{-9}$. Thus, its contribution is subleading to the corrections provided by $\cL_{\rm D4}$. Fourth, already the leading corrections break the degeneracy $h(r) = f(r)$ present in the Schwarzschild geometry. This may open up new ways to test the modified dynamics. In this context, we note that observations in the strong gravity regime essentially probe the geometry at $r = 3 G M$. The corrections then manifest themselves in the form
\be\label{bound-analysis}
\left. h(r) - h^{\rm SS}(r) \right|_{r = 3 G M} \propto \frac{M^4_{\rm Pl}}{M^4} \left[-4 \, \eta_2 + 2 \eta_5 - \eta_9 \right] \, ,
\ee
and similarly for $f(r)$. Thus lighter objects have the perspective of putting stronger bounds on the higher-derivative couplings. 
Finally, the results obtained from employing the field redefinition are recovered by setting the corresponding couplings to zero. In particular, the leading corrections agree with the analysis tracking $\cL_{\rm D6}^{\rm GS}$ \cite{Alvarez:2023gfg,Daas:2023axu}.

%---------------------------------------------------------------------
\section{Discussion and Conclusions}
%---------------------------------------------------------------------
A profound consequence of treating gravity either as an effective field theory or quantum theory is that the gravitational dynamics is supplemented by higher-derivative interactions. Implementing locality, these are readily organized in terms of a derivative expansion where the contributions of the higher-order terms are suppressed by a mass-scale, presumably given by the Planck mass. Owed to the geometric nature of the gravitational interactions, this quickly leads to a proliferation of admissible interaction terms \cite{Fulling:1992vm}. A common practice to reduce this complexity uses a field redefinition of the spacetime metric which, in the pure gravity case, allows to eliminate most of the higher-order terms. While significantly reducing the technical complexity of the subsequent analysis, this may come at the expense of obscuring the effects of the modified gravitational dynamics, in particular when matter degrees of freedom are included. One simply shifts effects from the gravitational dynamics into (modified) non-minimal interactions between the matter fields and the curvature of spacetime.

In this work, we determined the corrections to the metric functions characterizing static, spherically symmetric spacetimes based on a derivative expansion \emph{without resorting to field redefinitions}. Structurally, we highlighted the terms which give rise to new, massive Yukawa-type contributions and lead to new free parameters in the solution space. 
%This effect is beyond the effective field theory setting though, since it involves new degrees of freedom. 
Second, we tracked the PPN corrections to the metric potential.
%also keeping track of terms non-linear in the couplings. 
Owed to special properties $\cL_{\rm D4}$, this hierarchy does not follow the pattern of the derivative expansion 
%and the corrections appear as follows
%
\be\label{derivative-expansion}
\begin{array}{|c|c|c|c|c|}
\hline
\Big. {\rm derivative \; \; expansion} & \qquad \cL_{\rm D2} \qquad & \qquad \cL_{\rm D6} \qquad & \qquad \cL_{\rm D4} \qquad & \qquad \cL_{\rm D8} \qquad \\ \hline
\Big. {\rm PPN-order} & 1 & 6 & 8 & 9 
\\ \hline
\end{array} \, . 
\ee

The expansion \eqref{fhexp} needs to be used with care though. While valid in the asymptotically flat region, it may not be sufficient
%be difficult to use them 
to infer global properties of spacetime, as, e.g., the existence of an event horizon, since these may be situated outside the radius of convergence of the series. These features need to be corroborated, e.g., by a numerical integration in order to ensure that the approximation can be trusted in this regime \cite{Daas:2022iid}. The asymptotic expansion may then provide initial conditions which ensure that the spacetime is asymptotic flat, despite the exponential instability exhibited by \eqref{linsol}.

A second caveat comes with the assumption that \eqref{Dn-exp} contains local operators only. Treating general relativity as a quantum theory at the effective field theory level shows that the massless nature of the graviton also induces non-local terms in the gravitational dynamics \cite{Donoghue:2017pgk}. Currently, little is known about how these terms enter into the expansion \eqref{fhexp}. Since they come with fixed numerical coefficients which are not related to a new coupling constant, it would be highly desirable to include these contributions as well, thereby pinpointing the leading deviations predicted by the quantum theory.

\subsection*{Acknowledgements}
We thank M.\ Becker and L.\ Bouninfante for useful discussions. We also thank B.\ L.\ Giacchini and I.\ Kolar for correspondence on the first version of the manuscript. The work of C.L.\ is supported by the scholarship Becas Chile ANID-PCHA/2020-72210073. 

%---------------------------------------------------------------------

% Bibliography


%---------------------------------------------------------------------
\end{document}